\documentclass[usegraphicx,usenatbib,useAMS]{mn2e}
\usepackage{amsmath}
\usepackage{amssymb}
\usepackage{color}
\usepackage{url}
\usepackage{threeparttable}

\newcommand{\lya}{\ifmmode\mathrm{Ly}\alpha\else{}Ly$\alpha$\fi}
\newcommand{\lyb}{\ifmmode\mathrm{Ly}\beta\else{}Ly$\beta$\fi}
\newcommand{\igm}{\ifmmode\mathrm{IGM}\else{}IGM\fi}
\newcommand{\lae}{\ifmmode\mathrm{LAE}\else{}LAE\fi}
\newcommand{\h}{\ifmmode\mathrm{H}\else{}H\fi}
\newcommand{\hi}{\ifmmode\mathrm{H\,{\scriptscriptstyle I}}\else{}H\,{\scriptsize I}\fi}
\newcommand{\hii}{\ifmmode\mathrm{H\,{\scriptscriptstyle II}}\else{}H\,{\scriptsize II}\fi}
\newcommand{\cmb}{\ifmmode\mathrm{CMB}\else{}CMB\fi}
\newcommand{\qso}{\ifmmode\mathrm{QSO}\else{}QSO\fi}
\newcommand{\eor}{\ifmmode\mathrm{EoR}\else{}EoR\fi}
\newcommand{\cmmc}{\textsc{\small 21CMMC}}
\newcommand{\cmfst}{\textsc{\small 21CMFAST}}
\newcommand{\sense}{{\small 21}\textsc{cmsense}}

\newcommand{\GM}{GM15}

\newcommand\lsim{\mathrel{\rlap{\lower4pt\hbox{\hskip1pt$\sim$}}
        \raise1pt\hbox{$<$}}}
\newcommand\gsim{\mathrel{\rlap{\lower4pt\hbox{\hskip1pt$\sim$}}
        \raise1pt\hbox{$>$}}}

\title[SKA optimal core baseline design and observing strategy]{Optimal core baseline design and observing strategy for probing the astrophysics of reionization with the SKA}

\author[B. Greig et al.] {Bradley~Greig$^{1}$\thanks{E-mail:~bradley.greig@sns.it}, Andrei~Mesinger$^{1}$ and L\'eon V. E. Koopmans$^{2}$ \\
$^1$Scuola Normale Superiore, Piazza dei Cavalieri 7, I-56126 Pisa, Italy \\
$^2$Kapteyn Astronomical Institute, University of Groningen, PO Box 800, 9700 AV Groningen, The Netherlands \\
}

\begin{document}
\maketitle \begin{abstract}
\noindent
With the first phase of the Square Kilometre Array (SKA1) entering into its final pre-construction phase, we investigate how best to maximise its scientific return. Specifically, we focus on the statistical measurement of the 21 cm power spectrum (PS) from the epoch of reionization (\eor{}) using the low frequency array, SKA1--low. To facilitate this investigation we use the recently developed MCMC based \eor{} analysis tool \cmmc{} (Greig \& Mesinger). In light of the recent 50 per cent cost reduction, we consider several different SKA core baseline designs, changing: (i) the number of antenna stations; (ii) the number of dipoles per station; and also (iii) the distribution of baseline lengths.  We find that a design with a reduced number of dipoles per core station (increased field of view and total number of core stations), together with shortened baselines, maximises the recovered \eor{} signal.
With this optimal baseline design, we investigate three observing strategies, analysing the trade-off between lowering the instrumental thermal noise against increasing the field of view. SKA1--low intends to perform a three tiered observing approach, including a deep 100 deg$^{2}$ at 1000 h, a medium-deep 1000 deg$^{2}$ at 100 h and a shallow 10\,000 deg$^{2}$ at 10 h survey.
We find that the three observing strategies result in comparable ($\lsim$ per cent) constraints on our EoR astrophysical parameters.  This is contrary to naive predictions based purely on the total signal-to-noise, thus highlighting the need to use EoR parameter constraints as a figure of merit, in order to maximise scientific returns with next generation interferometers.

\end{abstract} 
\begin{keywords}
galaxies: high-redshift -- intergalactic medium -- cosmology: theory -- dark ages, reionization, first stars -- diffuse radiation -- early Universe
\end{keywords}

\section{Introduction}

The gravitational growth and collapse of initial density perturbations into the first astrophysical objects (i.e.\ stars and galaxies) signifies the transition from the end of the cosmic dark ages and beginning of the epoch of reionization (\eor{}). Emitted ultraviolet radiation from the first sources escapes into the surrounding intergalactic medium (\igm{}) forming ionized bubbles (\hii{} regions) in the pervasive neutral hydrogen fog which had enshrouded the Universe since recombination. As more astrophysical structures form, these \hii{} regions grow and overlap eventually permeating the entire Universe.

Evidently, the \eor{} is rich in astrophysical information. It provides insight into the formation, growth and evolution of structure in the Universe, the nature of the first stars and galaxies and their impact on the physics of the \igm{}, such as its ionization state and temperature \citep[see e.g.][]{Barkana:2007p2929, Loeb:2013p2936, Zaroubi:2013p2976}. The most promising cosmological probe of the reionization epoch is the radio detection of the redshifted 21 cm spin-flip transition of neutral hydrogen \citep[see e.g.][]{Furlanetto:2006p209,Morales:2010p1274,Pritchard:2012p2958}.

In search of the first direct detection from the \eor{}, several dedicated radio interferometers have been constructed or are planned. First-generation 21 cm experiments such as the Low Frequency Array (LOFAR; \citealt{vanHaarlem:2013p200,Yatawatta:2013p2980})\footnote{http://www.lofar.org/}, the Murchison Wide Field Array (MWA; \citealt{Tingay:2013p2997})\footnote{http://www.mwatelescope.org/} and the Precision Array for Probing the Epoch of Reionization (PAPER; \citealt{Parsons:2010p3000})\footnote{http://eor.berkeley.edu/} have been operating for the past few years. In principle, some of these experiments might yield a detection of the \eor{} 21 cm power spectrum (PS) (e.g.\ a marginal detection of the advanced stages of the reionization epoch; \citealt{Chapman:2012p344,Mesinger:2014p244}). Even without a detection, these experiments enable substantial progress in the understanding of various systematics and the characterisation and mitigation of astrophysical foregrounds crucial for future experiments.
  
Proposed, second-generation experiments such as the Square Kilometre Array (SKA; \citealt{Mellema:2013p2975})\footnote{https://www.skatelescope.org} and the Hydrogen Epoch of Reionization Array (HERA; \citealt{Beardsley:2014p1529})\footnote{http://reionization.org}, with substantially increased collecting areas will almost certainly detect the \eor{} 21 cm PS. However, unlike HERA, whose configuration has been adopted specifically to maximise its sensitivity to the PS, the SKA core baseline design and configuration are yet to be finalised. Recently, in the first step to achieving this, the generalised design for the SKA was announced. For SKA1--low, the low frequency instrument for which the detection of the \eor{} with 21 cm radiation is amongst the primary science cases, this allows for 50 per cent of the original outlined design\footnote{https://www.skatelescope.org/wp-content/uploads/2014/11/SKA-TEL-SKO-0000002-AG-BD-DD-Rev01-SKA1\textunderscore System\textunderscore Baseline\textunderscore Design.pdf}, an approximate total of $\sim130\,000$ dipole antennae. It is therefore crucial to investigate the optimal distribution of these dipoles for maximising the available \eor{} science. 

Several authors have constructed theoretical models to predict interferometer sensitivities facilitating the exploration of various aspects of the design optimisation process \citep{Morales:2004p3739,McQuinn:2006p109,Parsons:2012p95,Beardsley:2013p1300,Mellema:2013p2975,Trott:2014p3722}. The overall design depends on many factors; the size and distribution of the collecting area, the available frequency coverage, beam size (field of view), tracked or drift scanning and balancing the cost of correlating large numbers of antenna elements (see \citealt{Mellema:2013p2975} for discussions specific to the SKA). \citet{Parsons:2012p95} explored the optimisation of an interferometer layout, discussing the trade-off between improving the raw sensitivity through maximising the number of short, redundant baselines (reducing the necessary number of correlators) for 21 cm PS observations relative to the minimum redundant configurations best suited for imaging and foreground characterisation. 

For SKA1--low, we cannot simply assume a maximumly redundant design to recover the highest sensitivity measurement of the PS as it is seeking both a statistical characterisation and tomographic imaging of the reionization epoch \citep{Mellema:2013p2975}. Furthermore, SKA1--low has additional science goals beyond the \eor{}\footnote{http://pos.sissa.it/cgi-bin/reader/conf.cgi?confid=215}, including observing \hi{} line absorption, pulsars, magnetised plasma both in the galaxy and IGM, radio recombination lines and extrasolar planets. Therefore, optimising the final SKA1--low operating layout requires a more detailed exploration to satisfy all science goals, beyond the scope of this work. 

In this work, we focus instead on how best to distribute the available dipole antennae specifically targeting the core of the assumed layout configuration. Reducing the total number of dipoles allows several scenarios to be explored. Either (i) the number of dipoles per core antenna station can be reduced (increasing the beam size) or (ii) the number of core stations can be reduced (retaining the original beam size). Additionally (iii) the average baseline length can be reduced producing a more compact design, mitigating the reduced sensitivity due to the loss of dipoles. However, a more compact design may reduce the angular resolution of the instrument which could impact on the ability to provide images of individual \hii{} regions in the reionization epoch.

Another important aspect to consider is the optimal observing strategy to maximise the detection of the 21 cm PS. The shape and amplitude of the PS varies across the reionization epoch and depends on the astrophysics of the reionization process. The total noise of a radio interferometer is dominated by the sample (cosmic) variance on large scales, while small-scale modes are affected by the intrinsic detector (thermal) noise. By increasing the total integration time of a single observed patch sky, the thermal noise contribution to the total noise PS can be reduced, increasing the sensitivity on small to intermediate scale modes, while the contribution from sample variance remains constant. Conversely, for $N$ observations of the sky (where the total on sky time is fixed and distributed evenly across the $N$ observations), the contribution from sample variance to the total noise PS can be reduced by $\sqrt{N}$, thereby increasing the sensitivity to large-scale modes\footnote{In this instance, while the thermal noise per observation will have increased, by combining the thermal noise contribution across the $N$ observations will provide a slight improvement to the small-scale sensitivity.}.

In order to explore the optimisation of SKA1--low, we utilise the Monte Carlo Markov Chain (MCMC) based \eor{} analysis tool \cmmc{}\footnote{http://homepage.sns.it/mesinger/21CMMC.html} \citep{Greig:2015p3675} (hereafter \GM{}). Designed specifically for swift recovery of astrophysical constraints on the \eor{} given any observed statistical measure of the cosmic 21 cm signal, its flexibility is perfectly suited for optimising any telescope design and observing strategy. Importantly, this is the first optimisation study to use the precision of recovered \eor{} model parameters obtained from a mock measurement of the 21 cm signal as a figure of merit, improving on previous approaches based solely on the total integrated signal to noise (S/N) of the measured PS.

The remainder of this paper is organised as follows. In Section~\ref{sec:summary}, we summarise the key ingredients of \cmmc{} to be used throughout our analysis, before exploring in Section~\ref{sec:OptDesign} how best to distribute the available antenna dipoles to maximise the sensitivity of SKA1--low to the PS. With our optimal design, in Section~\ref{sec:OptStrat} we investigate the optimal observing strategy which maximises the recovered accuracy of \eor{} astrophysical constraints. Finally, in Section~\ref{sec:conclusion} we summarise the key conclusions regarding the optimisation of SKA1--low and finish with our closing remarks. Throughout this work, we adopt the standard set of $\Lambda$CDM cosmological parameters: ($\Omega_{\rm m}$, $\Omega_{\rm \Lambda}$, $\Omega_{\rm b}$, $n$, $\sigma_{8}$, $H_{0}$) =(0.27, 0.73, 0.046, 0.96, 0.82, 70~${\rm km\, s^{-1} \, Mpc^{-1}}$), measured from the \textit{Wilkinson Microwave Anisotropy Probe} \citep{Bennett:2013p3136} which are consistent with the latest results from Planck \citep{PlanckCollaboration:2014p3099}.

\section{Methodology} \label{sec:summary}

In \GM{}, we developed \cmmc{}: an MCMC driver of the publicly available semi-numerical code \cmfst{} \citep{Mesinger:2007p122,Mesinger:2011p1123}.  We showcased how it can be used to recover Bayesian constraints on astrophysical EoR parameters, given a mock PS observation with current and upcoming interferometers. Here we focus on the optimisation of SKA1--low, using as a figure of merit the precision to which we can recover astrophysical EoR parameters from the 21 cm PS. In the remainder of this section we summarise the relevant components of \cmmc{} necessary for performing this work, and defer the reader to \GM{} for more detailed discussions.

\subsection{Modelling reionization} \label{sec:21CMMC_reion}

Within \cmmc{} reionization is modelled using \cmfst{}, where 3D realisations of the \igm{} density, velocity, source and ionization fields are generated by evolving and smoothing a cubic volume of the linear density field using the Zel'dovich approximation \citep{Zeldovich:1970p2023}. ionization fields are then estimated using an excursion-set approach \citep{Furlanetto:2004p123}, comparing the time-integrated number of ionizing photons to the number of baryons within spherical regions of decreasing radius, $R$. Any cell within the simulation volume is classified as fully ionized if, 
\begin{eqnarray}
\label{eq:ioncrit}
\zeta f_{\rm coll}(\boldsymbol{x},z,R,\bar{M}_{\rm min}) \geq 1,
\end{eqnarray}
where $\zeta$ is the ionization efficiency which describes the conversion of mass into ionizing photons (see Section~\ref{sec:Zeta}) and $f_{\rm coll}(\boldsymbol{x},z,R,\bar{M}_{\rm min})$ is the fraction of collapsed matter within a spherical radius $R$ residing within haloes larger than $\bar{M}_{\rm min}$ \citep{Press:1974p2031,Bond:1991p111,Lacey:1993p115,Sheth:1999p2053}. For cells not fully ionized (partial ionization) the ionized fraction is set to $\zeta f_{\rm coll}(\boldsymbol{x},z,R_{\rm cell},\bar{M}_{\rm min})$, where $R_{\rm cell}$ is the minimum smoothing scale of the cell.
 
In GM15, we showed that \cmmc{} can accommodate various parametrisations of EoR astrophysics.  For simplicity, here we use a popular three parameter model, consisting of:
(i) the ionizing efficiency of high-$z$ galaxies; (ii) the mean free path of ionizing photons; (iii) the minimum virial temperature hosting star-forming galaxies. Although overly simplistic (as galaxy and IGM properties are averaged over redshift and/or halo mass dependences), this model provides the flexibility to describe a broad range of \eor{} signals while providing a straightforward physical interpretation, which we now briefly summarise.

\subsubsection{ionizing efficiency, $\zeta$} \label{sec:Zeta}

The ionizing efficiency of high-$z$ galaxies (Eqn.~\ref{eq:ioncrit}) can be expressed as 
 \begin{eqnarray} \label{eq:Zeta}
 \zeta = 30\left(\frac{f_{\rm esc}}{0.3}\right)\left(\frac{f_{\star}}{0.05}\right)\left(\frac{N_{\gamma}}{4000}\right)
 \left(\frac{2}{1+n_{\rm rec}}\right)
 \end{eqnarray}
 where, $f_{\rm esc}$ is the fraction of ionizing photons escaping into the \igm{}, $f_{\star}$ is the fraction of galactic gas in stars,  $N_{\gamma}$ is the number of ionizing photons produced per baryon in stars and $n_{\rm rec}$ is the typical number of times a hydrogen atom recombines. Although several of these parameters are uncertain, we showcase some possible values above, resulting in our fiducial choice of $\zeta=30$ used for our mock observation. Here we explore the range $\zeta\in[5,100]$. For reference, we also show on the top axis of the corresponding panels the parameter $f_{\rm esc}$, converted from $\zeta$ using the fiducial choices shown in Eqn.~(\ref{eq:Zeta}).

 \subsubsection{Mean free path of ionizing photons within ionized regions, $R_{\rm mfp}$} 

 The propagation of escaping ionizing photons through the \igm{} depends strongly on the abundances and properties of the absorption systems (Lyman limit as well as more diffuse systems). Below the resolution limits of \eor{} simulations, these act as photon sinks roughly dictating a maximum physical scale to which \hii{} bubbles are capable of growing around ionizing galaxies. Typically in \eor{} modelling this is parametrised by a maximum horizon for the ionizing photons, $R_{\rm mfp}$, corresponding to the maximum filtering scale in excursion-set \eor{} models (e.g. \citealt{Furlanetto:2005p4326}). Physically, this is often regarded as the mean free path of ionizing photons within ionized regions \citep[e.g.][]{OMeara:2007p3360,Prochaska:2009p3339,Songaila:2010p3348,McQuinn:2011p3293}, though during the EoR this maximum horizon can be smaller than the instantaneous mean free path \citep{Sobacchi:2014p1157}. Motivated by models containing sub-grid physics to model recombinations within these small absorption systems \citep[e.g.][]{Sobacchi:2014p1157}, we adopt a fiducial range of $R_{\rm mfp}\in[5, 20]$~cMpc, and the value of $R_{\rm mfp}=15$~Mpc for our mock observation. 
  
\subsubsection{Minimum virial temperature of star-forming haloes, $T^{\rm min}_{\rm vir}$} \label{sec:Mthresh}

We define the minimum threshold for a halo hosting a star-forming galaxy in terms of its virial temperature, $T^{\rm min}_{\rm vir}$.  This minimum halo virial temperature is set by the requirement of efficient: gas accretion, cooling and retainment of supernovae outflows. For reference, the virial temperature can be related to the halo mass via, \citep[e.g.][]{Barkana:2001p1634}:
\begin{eqnarray}
M_{\rm min} &=& 10^{8} h^{-1} \left(\frac{\mu}{0.6}\right)^{-3/2}\left(\frac{\Omega_{\rm m}}{\Omega^{z}_{\rm m}}
\frac{\Delta_{\rm c}}{18\pi^{2}}\right)^{-1/2} \nonumber \\
& & \times \left(\frac{T_{\rm vir}}{1.98\times10^{4}~{\rm K}}\right)^{3/2}\left(\frac{1+z}{10}\right)^{-3/2}M_{\sun},
\end{eqnarray}
where $\mu$ is the mean molecular weight, $\Omega^{z}_{\rm m} = \Omega_{\rm m}(1+z)^{3}/[\Omega_{\rm m}(1+z)^{3} + \Omega_{\Lambda}]$, and $\Delta_{c} = 18\pi^{2} + 82d - 39d^{2}$ where $d = \Omega^{z}_{\rm m}-1$.
Here we allow this threshold temperature to vary within the range $T^{\rm min}_{\rm vir}\in[10^{4},2\times10^{5}]$~K, with the lower limit corresponding to the atomic cooling threshold and the upper limit corresponding roughly to the observed Lyman break galaxy candidates. We choose $T^{\rm min}_{\rm vir} = 30\,000$ K (${\rm log\,}{T^{\rm min}_{\rm vir}} = 4.48$) for our mock observation, resulting in a reionization history which matches recent estimates based on the CMB optical depth \citep{Collaboration:2p3673}.

\subsection{Telescope noise profiles} \label{sec:TNP}

In order to explore the optimal core baseline design and observing strategy for SKA1--low, we compute the theoretical noise PS using the publicly available \textsc{python} module \sense{}\footnote{https://github.com/jpober/21cmSense}\citep{Pober:2013p41,Pober:2014p35}. For further details we defer the reader to  \citet{Parsons:2012p95} and \citet{Pober:2013p41,Pober:2014p35}, and instead briefly summarise the important assumptions.

The thermal noise PS is computed at each $uv$-cell  according to the following \citep[e.g.][]{Morales:2005p1474,McQuinn:2006p109,Pober:2014p35},
\begin{eqnarray} \label{eq:NoisePS}
\Delta^{2}_{\rm N}(k) \approx X^{2}Y\frac{k^{3}}{2\upi^{2}}\frac{\Omega^{\prime}}{2t}T^{2}_{\rm sys},
\end{eqnarray} 
where $X^{2}Y$ is a cosmological conversion factor between observing bandwidth, frequency and comoving volume, $\Omega^{\prime}$ is a beam-dependent factor derived in \citet{Parsons:2014p781}, $t$ is the total time spent by all baselines within a particular $k$ mode and $T_{\rm sys}$ is the system temperature, the sum of the receiver temperature, $T_{\rm rec}$, and the sky temperature $T_{\rm sky}$. 

Of great concern for 21 cm experiments is the estimation and removal of the bright foreground emission. Throughout this work, we conservatively choose to adopt the `moderate' foreground removal strategy of \citet{Pober:2014p35}. We excise Fourier modes which are localised into the foreground `wedge' within cylindrical 2D $k$-space \citep{Datta:2010p2792,Morales:2012p2828,Parsons:2012p2833,Trott:2012p2834,Vedantham:2012p2801,Thyagarajan:2013p2851,Liu:2014p3465,Liu:2014p3466}\footnote{It is important to note that this observing window is still not completely free of foregrounds. Fourier transforming the data cubes introduces side-lobe noise that can leak into this observing window and hence foreground removal is still required. However, conveniently the majority of this noise still occurs within this foreground `wedge'.}. To model the boundary separating the foreground `wedge' from the observing window, we fix its location to extend $\Delta k_{\parallel} = 0.1 \,h$~Mpc$^{-1}$ beyond the horizon limit \citep{Pober:2014p35}. This foreground removal strategy additionally includes the summation over instantaneously redundant baselines\footnote{Note, all 21 cm experiments contain redundant baselines, however, these are not always instantaneously redundant.} which can be coherently added to reduce the uncertainties in the thermal noise estimates \citep{Parsons:2012p95}. 

Within \sense{} the total noise PS is then computed by performing an inverse-weighted summation over all individual modes \citep[e.g.][]{McQuinn:2006p109}, combining both the contribution from the sample variance and the thermal noise (equation~\ref{eq:NoisePS}).
\begin{eqnarray} \label{eq:T+S}
\delta\Delta^{2}_{\rm T+S}(k) = \left(\sum_{i}\frac{1}{(\Delta^{2}_{{\rm N},i}(k) + \Delta^{2}_{21}(k))^{2}}\right)^{-1/2},
\end{eqnarray}
where $\delta\Delta^{2}_{\rm T+S}(k)$ is the total uncertainty from thermal noise and sample variance in a given $k$-mode and $\Delta^{2}_{21}(k)$ is the cosmological 21 cm PS from \cmfst{}. Here, we assume Gaussian errors for the cosmic-variance term, which on large scales is a good approximation. As in \GM{}, in addition to the aforementioned foreground `wedge', we include another level of conservatism to our errors, applying a strict $k$-mode cut, $|k| < 0.15$~Mpc$^{-1}$. Due to the location of the foreground `wedge' there is limited information available below this $k$-mode, however, modes above can still be affected by the foreground `wedge' due to its spherically asymmetric structure, hence we choose to apply both cuts.

Previously, both \citet{Pober:2014p35} and \GM{} assumed a drift scan observing mode for the SKA when estimating recovered \eor{} parameter constraints to aid a like-for-like comparison between the various 21 cm experiments. However, SKA1--low, is designed to operate in track scanning mode. In order to accurately discuss the optimisation of SKA1--low, we therefore adopt tracked scanning into our analysis. The major difference between the two scanning modes is on which spatial scales they are able to maximise the sensitivity. Drift scanning experiments are only able to observe chosen regions of the sky with low system temperatures for short periods of time (the actual time is dependent on the beam size) owing to the Earth's rotation. The short integration times result in more thermal noise,  although they allow several independent patches of sky to be observed per observing run, reducing the sample variance. The opposite is true for tracked scanning. As a result, a drift scanning mode has improved sensitivity on larger spatial scales (smaller $k$) whereas a tracked scan is favourable on moderate to smaller spatial scales.

Throughout this work we assume for SKA1--low that we are able to perform a single 6~h tracked scan on a patch of sky per night. This is a conservative choice to ensure our tracked regions of sky remain well above the horizon for the duration of the integration, mitigating any attenuation which may occur from observing close to the horizon. For all observations (i.e.\ our redshift range) we assume a bandwidth of 8~MHz and model the frequency dependent sky temperature as $T_{\rm sky} = 60\left(\frac{\nu}{300~{\rm MHz}}\right)^{-2.55}~{\rm K}$ \citep{Thompson2007} while adopting $T_{\rm rec} = 0.1T_{\rm sky} + 40~{\rm K}$ for the antenna dipoles.

\subsection{Recovering astrophysical constraints with \cmmc{}}

The large instantaneous bandwidth coverage available with the SKA-low (e.g.\ 50--350~MHz) should facilitate combining multiple epoch observations of the 21 cm PS to improve the overall constraining power. The improvements arise from the effective sampling of the redshift evolution in the 21 cm PS on large spatial scales, which are most sensitive to the \eor{} physics (see Fig.~2 of \GM{}). 

Our mock observation is generated using a 500~Mpc, 756$^3$ simulation, assuming $\zeta = 30$, $R_{\rm mfp} = 15$~cMpc and $T^{\rm min}_{\rm vir} = 3\times10^{4}$~K. All astrophysical constraints are recovered assuming the linear combination of the individual likelihoods of the PS observations at $z=8$, 9 and 10. By assuming an 8~MHz bandwidth, which corresponds to a $\Delta z < 1$, we ensure there is no overlap in redshift space for these observations.

\section{Optimal Core Baseline Design} \label{sec:OptDesign}

\begin{table*}
\begin{tabular}{@{}lccccccc}
\hline
Design & Diameter & Single beam FoV & \# Antenna & Collecting area & Description\\
layout & (m) & (deg$^{2}$ @ 150~MHz) &  stations &  (m$^2$) & \\
\hline
\vspace{0.8mm}
\#1 & 35.0 & 14.2 & 433 & 416\,595 & Original SKA design, number of stations halved\\
\vspace{0.8mm}
\#2 & 24.8 & 28.4 & 866 & 416\,595 & Original SKA design, dipoles per station halved\\
\vspace{0.8mm}
\#3 & 35.0 & 14.2 & 433 & 416\,595 & Baselines reduced by $\sqrt{2}$, number of stations halved\\
\vspace{0.8mm}
\#4 & 24.8 & 28.4 & 866 & 416\,595 & Baselines reduced by $\sqrt{2}$, dipoles per station halved\\
\hline
\end{tabular}
\caption{Summary of the different design layouts we consider for SKA1--low. The recently announced general design calls for a 50 per cent reduction of the originally planned SKA1--low, therefore we consider various ways of distributing such a reduction. We compute the single beam field of view (FoV) using FoV$_{\rm beam}$ = $\frac{\upi}{4}\left(\frac{1.3\lambda}{D_{\rm station}}\right)^{2}$.}
\label{tab:Designs}
\end{table*}  

The original design brief for SKA1--low core consisted of $\sim$250\,000 dipole antenna distributed across 866 core\footnote{Here, we refer only to the core of SKA1--low. A further 45 stations were to be configured into spiral arms extending outward as far as 50~km, to allow maximum baseline lengths of $\sim$100~km. We do not consider these stations as they add very little to the sensitivity of the 21 cm PS. Instead these aid calibration and foreground removal \citep{Mellema:2013p2975}.} antenna stations. Each station is 35~metres in diameter and arranged into a densely configured compact core containing $\sim$75 per cent of the antenna stations out to $\sim$1~km. Depending on the observing frequency each station would have a field of view of $\sim$20~deg$^{2}$. With a 50 per cent reduction to the original SKA1--low design, the question is, how best can these antenna dipoles be distributed to maximise the scientific returns for the \eor{}?

There are several options which can be considered.  To reduce cost one can either: (i) cut the total number of core antenna stations; or (ii) cut the number of dipoles per station (subsequently increasing the field of view).  Moreover for both (i) and (ii) one could also reduce the average baseline lengths for the core antenna stations\footnote{Note, we consider only reducing the average baseline lengths of the core antenna stations, baselines beyond $\sim$10 km used for calibration and foregrounds would remain in their original configuration.}, increasing the sensitivity over a narrower $k$-range. The reduction in the spatial resolution which comes with the compactification of the core, does not impact our EoR parameter constraints, as the bulk of the constraining power comes from larger scales.  Nevertheless, the loss in spatial resolution would adversely impact the proposed direct imaging of the EoR.  It is still not quantifiably clear what we will learn from such images. A more detailed exploration, beyond the scope of this present work, is required to quantify how an image-based EoR analysis is impacted by the choice of baseline lengths.
 
\subsection{SKA layouts} \label{sec:OptLayout}

Prior to exploring these options, we must first describe the layout with which we choose to model SKA1--low. It is planned  that SKA1--low will approximate a circularly symmetric array whose stations are randomly distributed and fall off with radius approximating a Gaussian distribution. At the same time, it should be as compact as possible, with the central core requiring a tightly packed configuration with a filling factor as close to unity as possible. The tightest packed configuration for circular stations is the hexagonal configuration. Indeed, \citet{Parsons:2012p95} showed that a hexagonal configuration provides the greatest level of maximum baseline redundancy increasing the overall telescope sensitivity to these Fourier modes, which in turn are best suited for the statistical recovery of the 21 cm PS.

Although this configuration is by no means the final baseline design, for the remainder of this work, we assume that SKA1--low has a central, densely packed core in a hexagonal configuration out to a certain radius, beyond which the remaining stations are randomly distributed falling off with radius (neglecting stations in the spiral arms for calibration and foreground removal). The radius of the hexagonal core is not fixed, but chosen per design to approximate as close as possible the required distribution of stations with radius and maintaining a filling factor as close to unity as possible, in general accordance with the SKA baseline design. We shall consider four distinct designs for the distribution of the dipole antennas, described below and summarised in Table~\ref{tab:Designs}.

\begin{figure*} 
	\begin{center}
		\includegraphics[trim = 0.8cm 0.5cm 0cm 0.3cm, scale = 0.88]{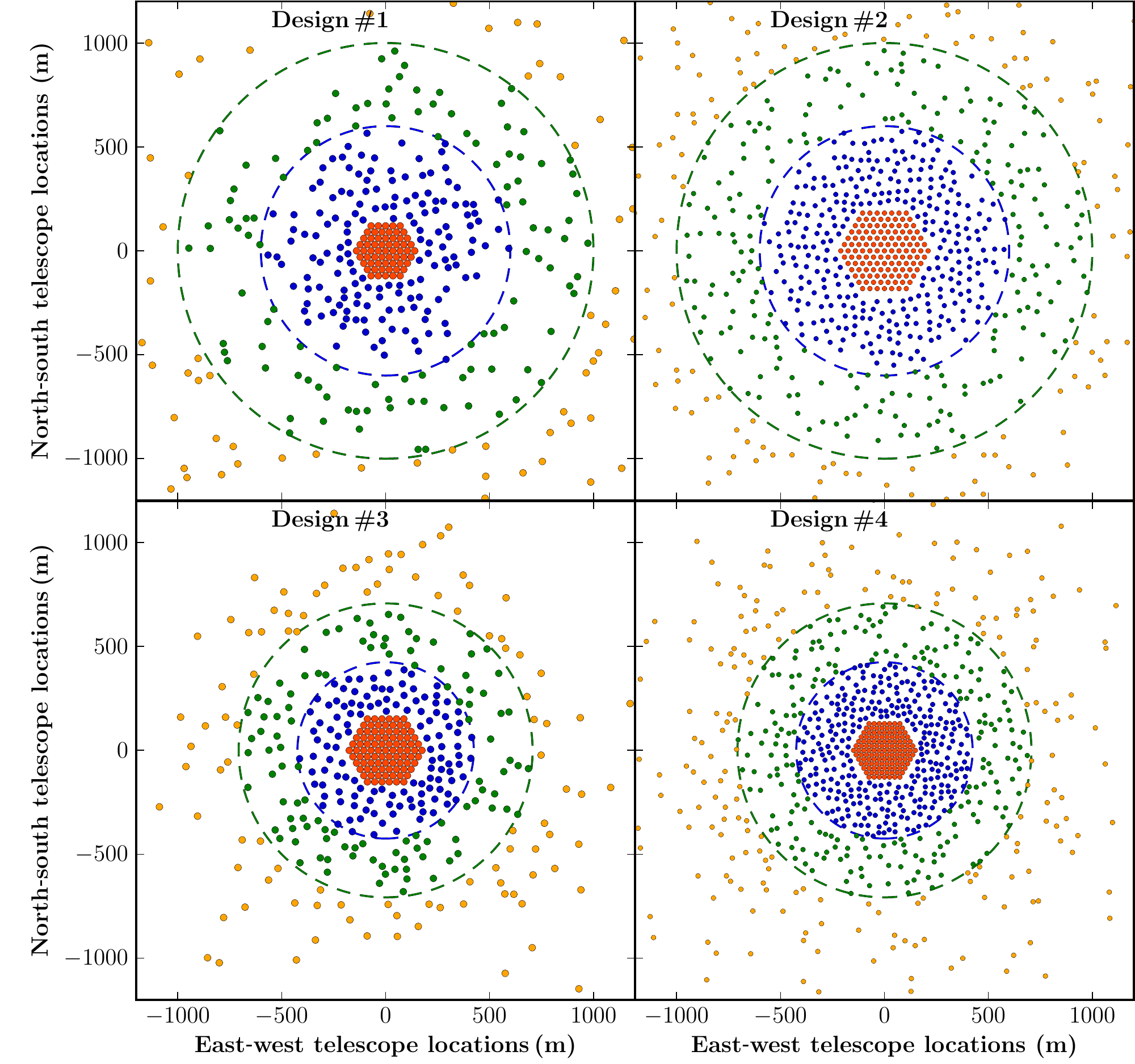}
	\end{center}
\caption[]
{A schematic of the four core baseline designs we consider for SKA1--low. Design \#1 (top left): 433 core antenna stations with a diameter of 35 m. Design \#2 (top right): 866 core antenna stations with half the number of dipoles per station as design \#1 (station diameter 35/$\sqrt{2}$ m). Design \#3 (bottom left): Same as design \#1 except reducing the baseline length by $\sqrt{2}$. Design \#4 (bottom right): Same as design \#2 except reducing the baseline length by $\sqrt{2}$. The relative physical size of the filled circles correspond to the true diameter of the antenna stations. For all designs the core antenna stations are normally distributed with radius, including a central densely packed `inner' core in a hexagonal configuration (red circles). For design \#1 and \#2, blue, green and yellow circles denote stations within a 600~m radii (blue dashed curve), 1000~m radii (green dashed curve) and stations beyond 1000~m from the core respectively. For design \#3 and \#4, these radii are reduced by $\sqrt{2}$. }
\label{fig:SKA_Layout}
\end{figure*}

\begin{itemize}
\item Design \#1: Original design for SKA1--low, halving the number of antenna stations down to a total of 433 (retaining the original number of dipoles per station, i.e.\ the original field of view). This design retains 75 per cent of the stations within a radius of $\sim$1~km, and contains 61 stations in a hexagonal configuration in the inner core. 
\\
\item Design \#2: Original design for SKA1--low, halving the number of dipole antennas per station. We retain the originally planned 866 stations, with a reduction in the station diameter of $\sqrt{2}$ (doubling the field of view of the antenna stations). To construct this design, we originally assume an antenna station diameter of 35~m before shrinking the station diameter. A total of 127 stations are included in the hexagonal inner core, and 75 per cent of the stations are within a radius of $\sim$1~km.
\\
\item Design \#3: Same as design \#1, except we additionally reduce the average baseline length by $\sqrt{2}$ producing a more compact overall design. With this, we increase the size of the hexagonal inner core to 91 stations, and now contain 75 per cent of the stations within a radius of $\sim$700~m.
\\
\item Design \#4: Same as Design \#2, reducing the average baseline length by $\sqrt{2}$ to produce a more compact overall design. With the increased number of stations in this design (866), the inner hexagonal core retains the number of stations as design \#2 but is shrunk to be fully compact. This design maintains 75 per cent of the stations within a radius of $\sim$700~m.
\end{itemize}

\begin{figure} 
	\begin{center}
		\includegraphics[trim = 0cm 0.7cm 0cm 0.3cm, scale = 0.43]{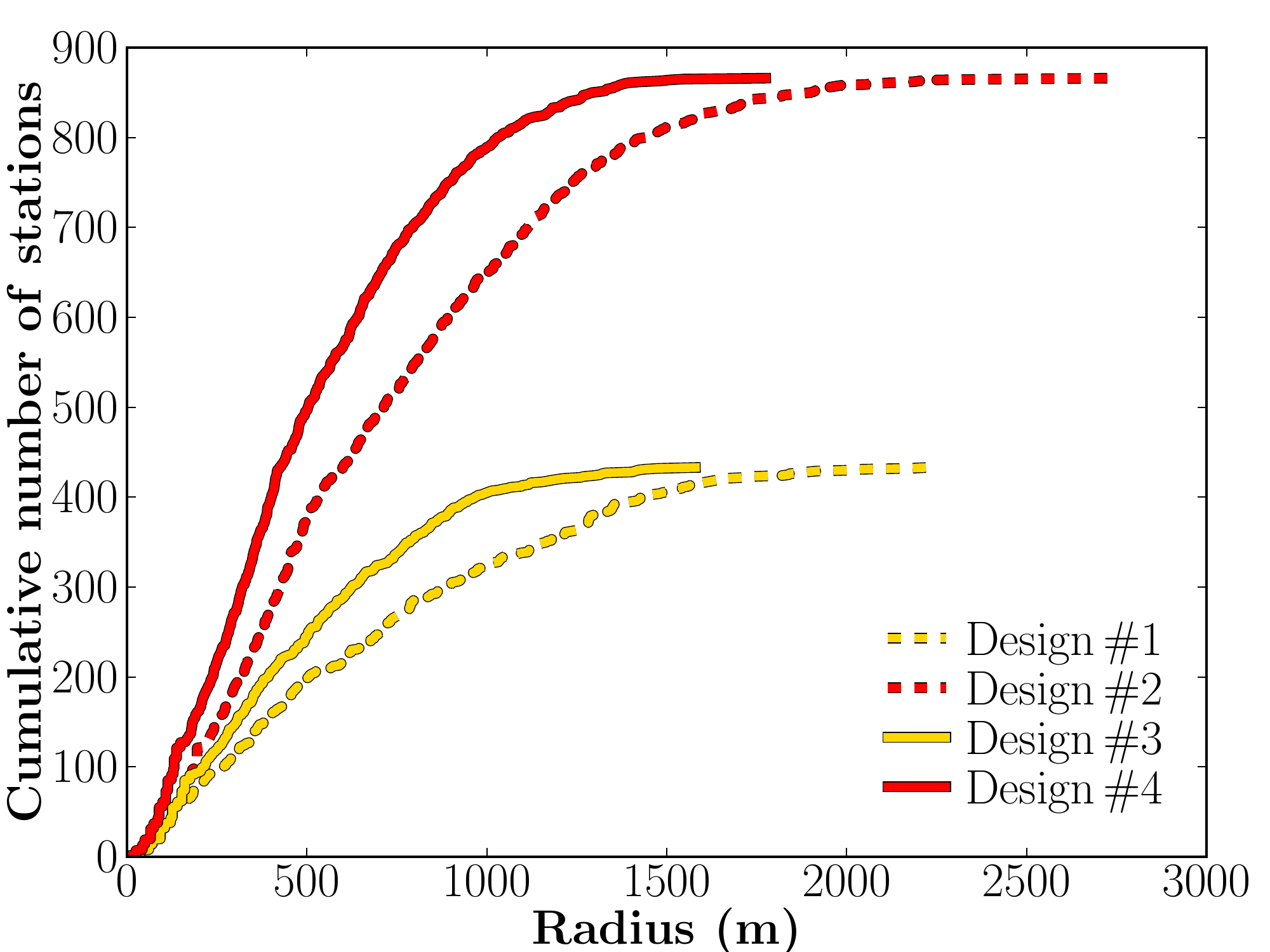}
	\end{center}
\caption[]
{The cumulative distribution of SKA1--low telescope antenna as a function of radius for all instrument designs we consider. The red curves correspond to retaining the original number of core antenna stations by halving the number of dipoles per core antenna station, while the yellow curves correspond to halving the number of core stations. Dashed curves correspond to the original baseline lengths while the solid curves denote designs where the average baseline length is reduced by $\sqrt{2}$.}
\label{fig:SKA_CDF}
\end{figure}

For a visual description of these designs we provide Fig.~\ref{fig:SKA_Layout}. All filled circles correspond to the relative diameter of the antenna stations. Within the inner hexagonal core stations are denoted by red circles, while blue, green and yellow circles correspond to stations within certain radii (emphasised by the circular dashed lines). For designs \#1 and \#2, these correspond to stations within a 600~m radii, 1~km radii and stations outside a 1~km radii of the centre, respectively. For designs \#3 and \#4 (with reduced baselines), these radii are reduced by $\sqrt{2}$. In Fig.~\ref{fig:SKA_CDF}, we additionally provide the cumulative number of antenna stations with radius to better highlight the distribution of antenna stations amongst our four chosen designs.

\subsection{Power spectrum sensitivity per beam}

\begin{figure*} 
	\begin{center}
		\includegraphics[trim = 0.2cm 0.35cm 0cm 0.4cm, scale = 1.1]{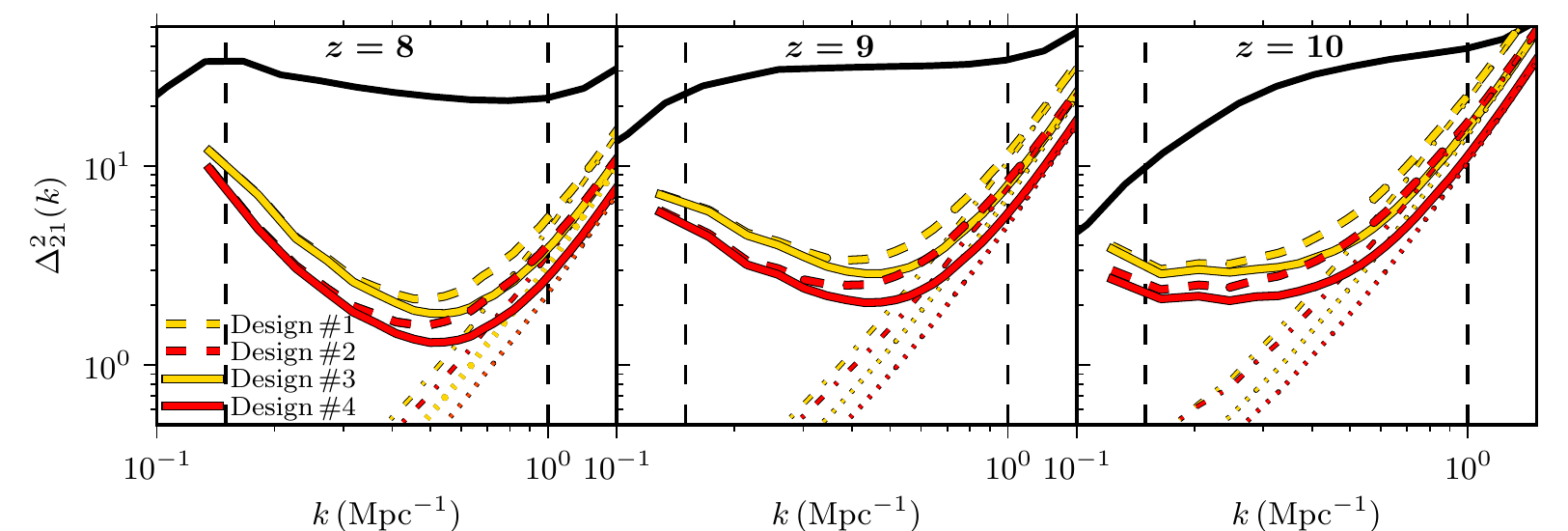}
	\end{center}
\caption[]
{The estimated total noise PS for a single beam pointing of 1000~h for each baseline design we consider for SKA1--low at three different observational epochs ($z=8$, 9 and 10). The solid black curve corresponds to our fiducial reionization model, with ($\zeta$, $R_{\rm mfp}$, $T^{\rm min}_{\rm vir}$) = (30, 15~Mpc, $3\times10^{4}$~K) with a corresponding neutral fraction of $\bar{n}_{\hi{}}$ = 0.5, 0.7 and 0.8, respectively. Red and yellow curves denote designs whereby we either halve the number of dipoles per core antenna station or the total number of antenna stations, respectively. Dashed curves denote the retention of the original baseline lengths, whereas the solid curves correspond to reducing the average baseline length by $\sqrt{2}$ producing a more compact instrument. Dot-dashed and dotted curves indicate the thermal noise contribution only for the dashed and solid curves, respectively. Vertical dashed lines at $k = 0.15$ and 1.0~Mpc$^{-1}$ demarcate the physical scales within which we fit the 21cm PS within \cmmc{}.}
\label{fig:Sensitivity_Design}
\end{figure*}

In Fig.~\ref{fig:Sensitivity_Design}, we compute the total noise PS (Eqn.~\ref{eq:T+S}) from \sense{} for each design assuming a single beam pointing integrated for a total of 1000~h. Here, and throughout the remainder of this work, we assume that SKA1--low can always be calibrated to reach its theoretical noise limit as defined by the total noise PS. We show the total theoretical noise PS for all designs at each redshift for our multiple epoch observing strategy ($z=8$, 9 and 10, left, centre and right panel, respectively), where the solid black curve corresponds to the mock 21 cm  PS. The red and yellow curves  denote halving the number of dipoles per station or halving the total number of stations, respectively. The dashed curves denote the retention of the original baseline distribution, whereas the solid curves denote reducing the average baseline lengths. Dot-dashed and dotted curves denote the contribution from only the thermal noise component for the dashed and solid curves respectively.

The major difference between the curves in Fig.~\ref{fig:Sensitivity_Design} stems from the increased field of view of designs \#2 and \#4, relative to the other designs. Reducing the number of dipoles per core antenna station, doubles the effective field of view (see Table~\ref{tab:Designs}), improving the sensitivity to Fourier modes most relevant for measuring the 21 cm PS. The larger effective sky area observed per beam pointing reduces the error contribution from sample variance, as well as that of the thermal noise (see Eqn.~\ref{eq:NoisePS}) since each mode in the sky can be observed with the same S/N per resolution element over a larger part of the sky. As a result, for the same total observing time designs \#2 and \#4 are optimal.

\subsection{Power spectrum sensitivity per fixed survey area}

\begin{figure*} 
	\begin{center}
		\includegraphics[trim = 0.2cm 0.35cm 0cm 0.4cm, scale = 1.1]{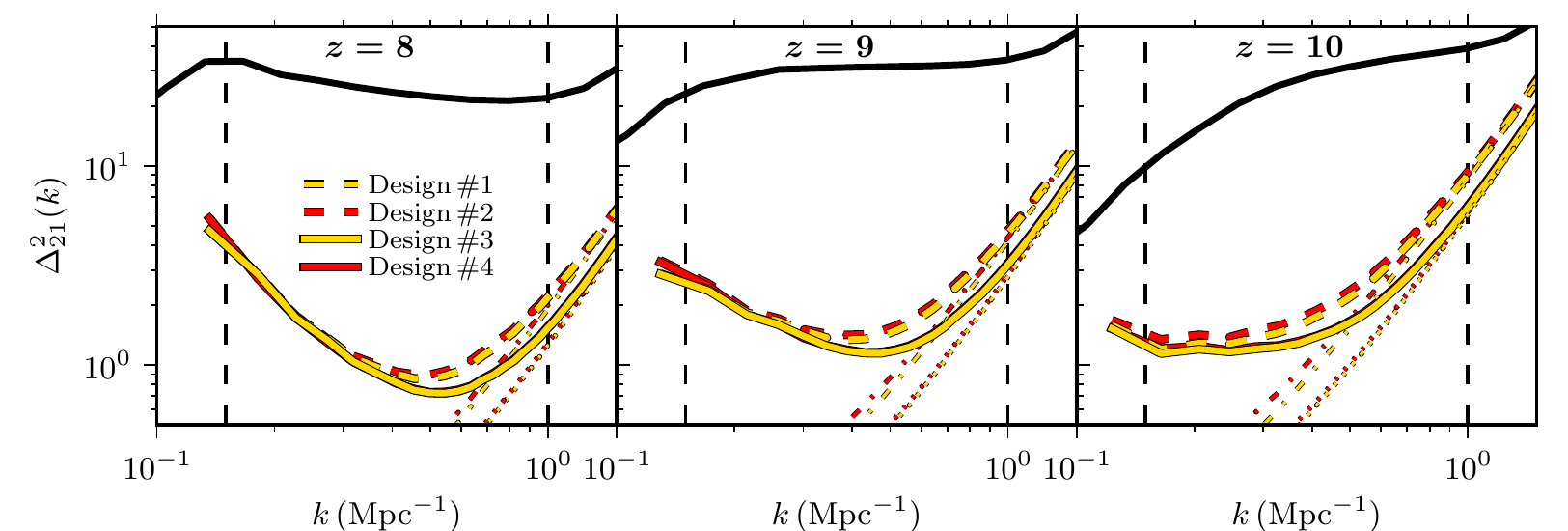}
	\end{center}
\caption[]
{Same as Fig.~\ref{fig:Sensitivity_Design}, except now we compare our four baseline designs for the SKA1--low for a survey with a total observed area of sky of 100 deg$^{2}$ at 1000~h. This, as expected, mitigates the differences between the two options of halving the number of stations or halving the dipoles per station and emphasises the improvement gained on moderate to small scales by reducing the baseline lengths. Note, the total time on sky for designs \#1 and \#3 is twice as long as for designs \#2 and \#4 due to the reduced field of view (see Table~\ref{tab:Designs}).}
\label{fig:Sensitivity_Design_scaled}
\end{figure*}

In order to highlight this, in Fig.~\ref{fig:Sensitivity_Design_scaled}, we provide the same as Fig.~\ref{fig:Sensitivity_Design}, except we normalise the sensitivity curves to the same total survey area footprint. Here, we consider a total area of 100 deg$^{2}$ on the sky observed at 1000~h per beam. Note for this survey, designs \#1 and \#3, with half the field of view, effectively take twice as long in terms of total time spent observing on the sky, due to the increased pointings required to obtain the fixed survey area. With the impact of field of view now normalised out, we find designs \#1 and \#2 to be equivalent, and the same for designs \#3 and \#4. This is theoretically expected as by observing the same total survey area, all designs have the same error contribution from sample variance. Additionally, Fig.~\ref{fig:Sensitivity_Design_scaled} clearly emphasises the impact of reducing the average baseline lengths to produce a more compact instrument. On large scales there is no improvement, as the total error is completely dominated by sample variance, however, on intermediate to smaller scales the impact becomes more pronounced. By reducing the average baseline length, in effect we increase the number of core antenna stations separated by baselines sensitive to these Fourier modes, improving the overall sensitivity of the instrument. However, if we were to continue down to even smaller scales we would eventually see a turn over, with decreased sensitivity for very small scale modes. 

Clearly, from Figs~\ref{fig:Sensitivity_Design} and~\ref{fig:Sensitivity_Design_scaled} the optimal design for SKA1--low is then design \#4, from which we benefit both from the reduction in baseline lengths, and the decreased total on sky observing time (due to a larger FoV). However, our choice of design \#4 over design \#3 does not factor in other considerations. For example, with double the number of core antenna stations than design \#3, design \#4 will have a substantially higher computational cost for correlating the signal. Additionally, construction costs will be larger owing to more stations being required. Despite these caveats, for the remainder of this work, we shall consider design \#4 to be the optimal design for the SKA1--low.
Moreover, we will only consider surveys of total fixed area and not total observing time. Therefore, any conclusions drawn would be equivalent between designs \#3 and \#4. However, if for example the observing time was additionally fixed (owing to survey restrictions), design \#4 would outperform design \#3 by effectively being able to spend twice as long per beam pointing, providing a gain in S/N over the entire survey area ($\sqrt{2}$ for imaging and factor of two for the PS sensitivity).

\section{Optimal Observing Strategy and EoR parameter forecasts} \label{sec:OptStrat}

With our optimal design for SKA1--low, we now seek to determine the optimal observing strategy. The overall sensitivity of a radio interferometer to the 21 cm PS is dictated by the relative contributions from the sample (cosmic) variance and thermal noise. On large scales, the noise PS is dominated by the contribution from sample variance which scales as $\Delta^2_{SV} \propto \Omega_{\rm A}^{-1/2}$ (for an assumed fixed total collecting area of the instrument; e.g.~\citealt{Mellema:2013p2975}), where $\Omega_{\rm A}$ is the total survey area. On small scales, the noise PS is instead dominated by thermal noise and scales as $\Delta^2_{N} \propto \Omega_{\rm A}^{-1/2} / t$, where $t$ is the time spent on each pointing. To maximise the EoR science return, would a fixed amount of telescope time be better spent increasing $\Omega_{\rm A}$ or $t$?\footnote{In obtaining these scaling relations, we have adopted our optimal SKA design, design \#4, which has a fixed field of view. Therefore to observe the requisite total sky area we assume it can always be obtained by tiling the survey by $N$ ($\sim \Omega_{\rm A}/\Omega_{\rm FoV}$) observations of the antenna station field of view for $t$ hours.} 

SKA1--low anticipates performing a three tiered observational strategy in order to balance between the competing noise contributions, resulting in a relatively uniform S/N coverage over a broad stretch of $k$-modes. These include a deep, medium-deep and a shallow survey \citep{Koopmans:2015:p1} which we outline below;

\begin{itemize}
\item 100 deg$^{2}$ at 1000~h (deep): a single, deep integration of a 100 deg$^{2}$ patch of sky observed at 1000~h. Such a survey is designed for the imaging of the reionization epoch, seeking the characterisation and detection of individual \hii{} regions. This approach minimises the thermal noise contribution to the total noise PS, favouring intermediate to small-scales of the 21 cm PS at the expense of an increasing sample variance component reducing the sensitivity to the large-scale modes in the PS from this survey (note however that the thermal noise as shown in Fig.~\ref{fig:Sensitivity_Design} and~\ref{fig:Sensitivity_Design_scaled} is smallest on the largest angular scales; see e.g. \citealt{Koopmans:2015:p1}).  This survey would also extend furthest in redshift, allowing us to probe the pre-reionization epoch.
\\
\item 1000 deg$^{2}$ at 100~h (medium-deep): an intermediate observing strategy, balancing sample variance and thermal noise.
\\
\item 10\,000 deg$^{2}$ at 10~h (shallow): a short integration time per pointing\footnote{Note, 10~hrs with SKA1--low is roughly equivalent to 600 hrs with LOFAR, which aims for a statistical detection of the \eor{} signal \citep{Harker:2010p4317}. Therefore, a single 10 hr pointing with SKA1-low
  will be equivalent to the deepest integrations with LOFAR to date.} to build a wide/shallow survey at 10\,000 deg$^{2}$ after $\sim400$ pointings. This survey substantially reduces the sample variance contribution, focussing the sensitivity of SKA1--low to the largest-scale modes at the expense of an increasing thermal noise contribution on all other scales.
\end{itemize}

\begin{figure*} 
	\begin{center}
		\includegraphics[trim = 0.2cm 0.35cm 0cm 0.4cm, scale = 1.1]{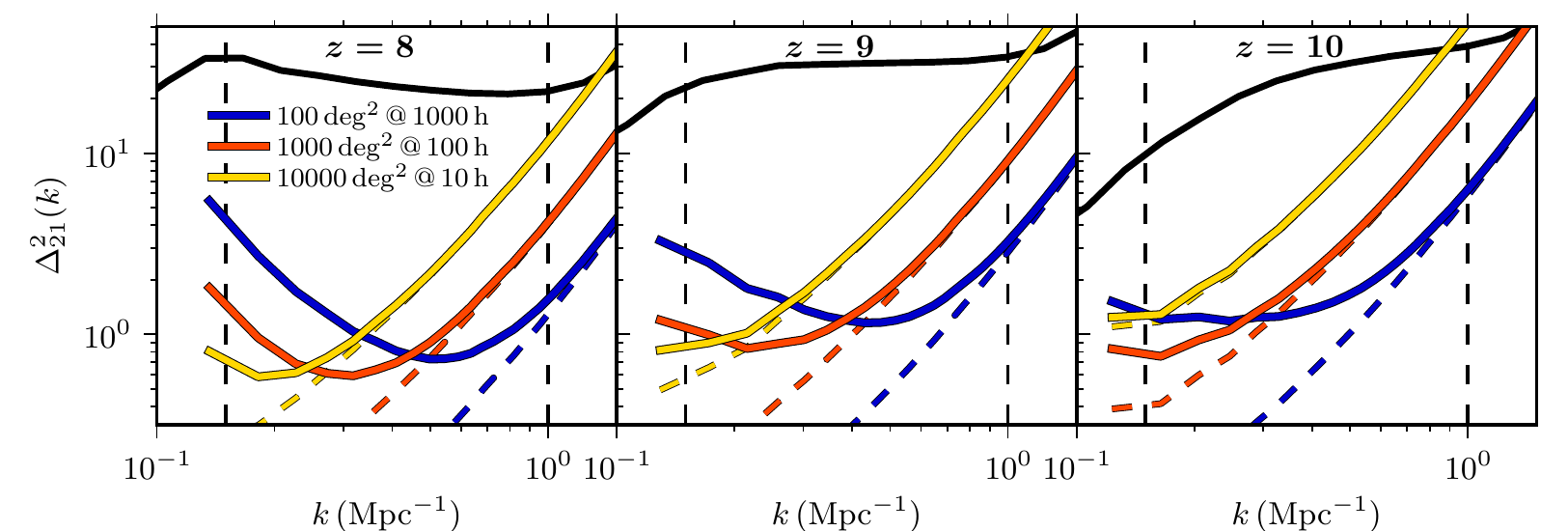}
	\end{center}
\caption[]
{The estimated total noise PS for the three \eor{} observing strategies to be performed by SKA1--low. A single, deep 1000 hr integration over a 100 deg$^{2}$ survey footprint (blue curve), an intermediate survey area of 1000 deg$^{2}$ at 100 h (red curve) and many, short 10 h integrations to produce a wide/shallow 10\,000 deg$^{2}$ (yellow curve). Dashed curves correspond to the contribution from the thermal noise PS only. While shortening the integration time of an individual patch of the sky increases the thermal noise resulting in a reduction of the sensitivity on intermediate to small scales, it reduces the sample variance contribution to the total noise improving the sensitivity on larger physical scales. }
\label{fig:Sensitivity_Strat}
\end{figure*}

It is important to note that we do not seek to choose only one of the above observing strategies as SKA1--low aims to perform all three in order to obtain a uniform PS S/N coverage over the widest possible $k$-range. Instead, we provide forecasts for how well each observing strategy will recover the astrophysical parameters of the EoR.

In Fig.~\ref{fig:Sensitivity_Strat} we present the total noise PS for each of the three observing strategies we consider for the same three redshift observations ($z=8$, 9 and 10). Again, the solid black curve denotes the theoretical 21 cm PS for our fiducial \eor{} model. The blue, red and yellow curves correspond to the three observing strategies, 100 deg$^{2}$ at 1000~h, 1000 deg$^{2}$ at 100~h and 10\,000 deg$^{2}$ at 10~h, respectively, while the dashed curves correspond to the thermal noise contribution. This figure clearly shows  how each observing strategy benefits from their respective choice of minimising the thermal noise or sample variance contribution. Across all three panels, we observe a shift in the minima of the total noise PS to larger spatial scales (small $k$) as the sample variance contribution is decreased by shortening the observing time per patch of sky.\footnote{Note that at $z=10$, and to a lesser extent at $z=9$, the total noise PS for our shallow observing strategy is not able to reach its minima due to the assumed contamination by foregrounds. As a result, this will impact on the total available constraining power from this strategy as we discuss below.} The converse is true on small scales due to the increasing thermal noise contribution.  Together, the three observing strategies provide relatively uniform S/N coverage over this range of $k$.

The total integrated S/N of the 21 cm PS (which is \eor{} model dependent), summed across all Fourier modes, is relatively similar across the three observing strategies, with the 1000~h observing strategy yielding the highest total sensitivity.
However, the differences between EoR models are greatest on the largest scales shown in Fig. 2 of \GM{}; we therefore expect that the integrated S/N is not a good proxy for how well EoR astrophysics can be constrained.

\subsection{Theoretical limit on recovered \eor{} parameters} \label{sec:woModUncert}

As discussed in \GM{}, even if our EoR parametrisation is correct, modelling errors are introduced when simulating the EoR with \cmfst\ (compared with an idealised, infinite resolution, first-principle numerical simulation).  Characterisation of these uncertainties over a broad parameter range (e.g. by calibrating to more detailed radiative transfer simulations) will be essential to maximising the achievable scientific gains from the SKA. Here we begin by forecasting the recovered EoR parameters without including any additional modelling uncertainty, before relaxing this assumption in Section~\ref{sec:wModUncert}.

\begin{figure*} 
	\begin{center}
		\includegraphics[trim = 0.2cm 0.5cm 0cm 0.5cm, scale = 0.88]{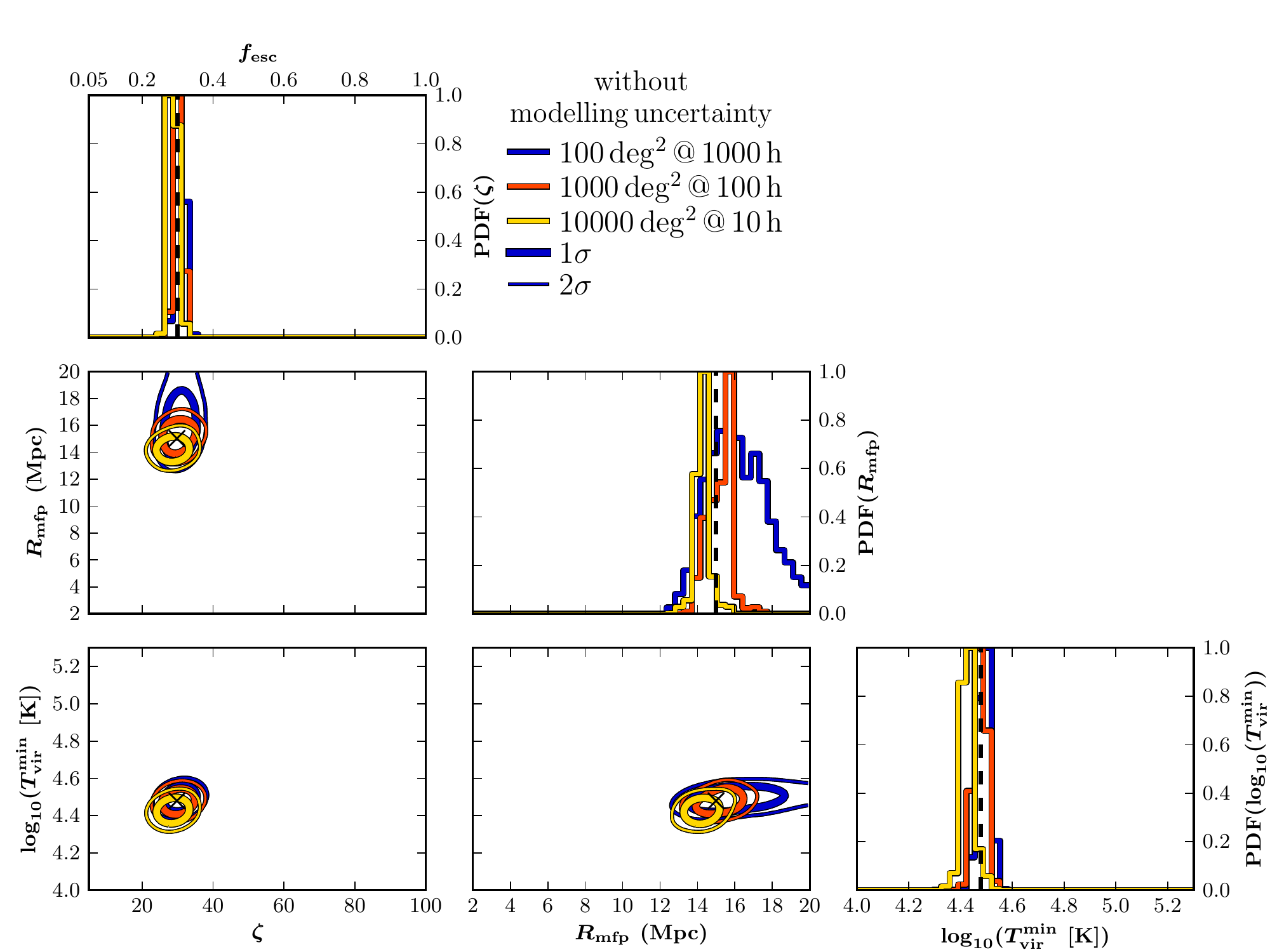}
	\end{center}
\caption[]
{The recovered constraints on our reionization model parameters for three different observing strategies, combining three independent ($z=8, 9$~and~10) observations of the 21 cm PS. We consider a single, deep 100 deg$^{2}$ observation at 1000~h (blue), an intermediate medium-deep 1000 deg$^{2}$ observation at 100~h (red) and a wide/shallow observation of 10\,000 deg$^{2}$ at 10~h (yellow). Across the diagonal panels we provide the 1D marginalised PDFs for each of our reionization parameters [$\zeta$, $R_{\rm mfp}$ and  log$_{10}$$(T^{\rm min}_{\rm vir})$ respectively] and highlight our fiducial model parameter by a vertical dashed line. Additionally, we provide the 1 (thick) and 2$\sigma$ (thin)  2D joint marginalised likelihood contours for our three reionization parameters in the lower left corner panels of the figure (crosses mark our fiducial reionization parameters).}
\label{fig:ObsStrat_woRT}
\end{figure*}

In Fig.~\ref{fig:ObsStrat_woRT} we provide both the 1 and 2D marginalised probability distribution functions (PDFs) recovered from \cmmc{} for our \eor{} model parameters for the three different observing strategies. In this figure, and for the remainder of this work we assume uniform (flat) priors on all recovered \eor{} parameters within their allowed ranges outlined in Section~\ref{sec:21CMMC_reion}. Diagonal panels show the recovered marginalised 1D PDFs for each corresponding \eor{} parameter with the vertical dashed line indicating the location of our fiducial input value. In the lower left corner panels, we provide the 1 and 2$\sigma$ 2D marginalised likelihood contours for our \eor{} model parameters, where crosses denote their fiducial values.

The confidence limits from this figure are quantitatively summarised in the top half of Table~\ref{tab:3param_strat}. We report the median recovered values, and the 16th and 84th percentiles. Assuming the recovered 1D marginalised PDFs for each \eor{} parameter can be modelled by a Gaussian distribution\footnote{By performing a MCMC recovery of the astrophysical \eor{} parameters, the recovered marginalised PDFs for each parameter need not be symmetric. However, to a good approximation we find the PDFs of our \eor{} parameters to closely resemble a Gaussian distribution.}, we find the fractional uncertainty to which we recover these to be 3.7/3.4/3.6 per cent for $\zeta$, 9.6/4.1/2.7 per cent for $R_{\rm mfp}$ and 0.5/0.5/0.6 per cent for log$_{10}$$(T^{\rm min}_{\rm vir})$ for our deep, medium-deep and shallow observing strategies, respectively.\footnote{It would be instructive to compare our recovered constraints from our this updated SKA1--low compared to the original, larger design considered in \GM{}. However, it is not straightforward to do so, as previously we only assumed a single beam pointing at 1000~h, relative to the fixed survey areas we consider in this work.}

All three surveys recover tight constraints on our \eor{} parameters, highlighting the strength of SKA1--low at detecting the 21 cm PS. However, the medium-deep and shallow surveys somewhat outperform the deep observing strategy. This, however, is not surprising given that the majority of the constraining power on the astrophysical \eor{} parameters arises from sampling the large-scale modes at multiple redshifts (\GM{}).  This is especially true for the mean free path parameter, which at $R_{\rm mfp}\gsim15$~Mpc does not strongly impact the reionization history, but does change the 21 cm large scale power (see Fig.~2 in \GM{}); this signature is missed by the deep survey, and thus the resulting contours on $R_{\rm mfp}$ are broader than those corresponding to the other surveys.

It is additionally worth noting that we are \textit{only} using the PS to compute the likelihood.  The 21 cm signal during the EoR is highly non-Gaussian. Thus higher order statistics, taking advantage of the increased small-scale sensitivity of the deep survey, could provide further information.  In principle, the \cmmc{} framework allows for arbitrary figures of merit; we defer such studies of alternative statistics to future work.

\begin{table*}
\begin{tabular}{@{}lccccccc}
\hline
Observing strategy & & Parameter & & & $\bar{x}_{\hi{}}$ &  \\
(with/without modelling uncertainty) & $\zeta$ & $R_{\rm mfp}$~(Mpc) & log$_{10}$$(T^{\rm min}_{\rm vir})$ & $z = 8$ & $z = 9$ & $z = 10$\\
\hline
\vspace{0.8mm}
100 deg$^{2}$ @ 1000 h (without) & $30.66\substack{+1.20 \\ -1.11}$ & $15.94\substack{+ 1.77\\ -1.42}$ & $4.49\substack{+0.02 \\ -0.02}$ & $0.48\substack{+ 0.01\\ -0.01}$ & $0.71\substack{+0.01 \\ -0.01}$ & $0.84\substack{+ 0.01\\ -0.01}$\\
\vspace{0.8mm}
1000 deg$^{2}$ @ 100 h (without) & $30.25\substack{+1.02 \\ -1.07}$ & $15.43\substack{+ 0.31\\ -1.03}$ & $4.48\substack{+ 0.02\\ -0.02}$ & $0.48\substack{+ 0.01\\ -0.01}$ & $0.70\substack{+ 0.01\\ -0.01}$ & $0.83\substack{+ 0.01\\ - 0.01}$ \\
\vspace{0.8mm}
10\,000 deg$^{2}$ @ 10 h (without) & 28.71$\substack{+0.96 \\ -0.82}$ & 14.22$\substack{+0.22 \\ -0.19}$ & 4.43$\substack{+0.02 \\ -0.02}$ & 0.47$\substack{+0.01 \\ -0.01}$ & 0.69$\substack{+0.01 \\ -0.01}$ & 0.82$\substack{+0.01 \\ -0.01}$\\
\hline
\vspace{0.8mm}
100 deg$^{2}$ @ 1000 h (with 10 per cent) & 30.68$\substack{+2.44 \\ -2.18}$ & 15.49$\substack{+2.21 \\ -1.94}$ & 4.49$\substack{+0.05 \\ -0.05}$ & 0.48$\substack{+0.02 \\ -0.02}$ & 0.71$\substack{+0.01 \\ -0.01}$ & 0.84$\substack{+0.01 \\ -0.01}$\\
\vspace{0.8mm}
1000 deg$^{2}$ @ 100 h (with 10 per cent) & 30.62$\substack{+2.68 \\ -2.33}$ & 15.12$\substack{+1.95 \\ -1.66}$ & 4.49$\substack{+0.06 \\ -0.06}$ & 0.49$\substack{+0.02 \\ -0.02}$ & 0.71$\substack{+0.01 \\ -0.01}$ & 0.84$\substack{+0.01 \\ -0.01}$\\
\vspace{0.8mm}
10\,000 deg$^{2}$ @ 10 h (with 10 per cent) & 30.70$\substack{+3.44 \\ -2.84}$ & 14.96$\substack{+2.05 \\ -1.69}$ & 4.49$\substack{+0.07 \\ -0.07}$ & 0.48$\substack{+ 0.02\\ -0.02}$ & 0.71$\substack{+0.02 \\ -0.02}$ & 0.84$\substack{+0.01 \\ -0.01}$\\
\hline
\vspace{0.8mm}
100 deg$^{2}$ @ 1000 h (with 25 per cent) & 31.68$\substack{+6.08 \\ -4.45}$ & 14.81$\substack{+2.90 \\ -3.04}$ & 4.51$\substack{+0.11 \\ -0.11}$ & 0.49$\substack{+0.04 \\ -0.04}$ & 0.71$\substack{+0.03 \\ -0.03}$ & 0.84$\substack{+0.02 \\ -0.02}$\\
\vspace{0.8mm}
1000 deg$^{2}$ @ 100 h (with 25 per cent) & 31.84$\substack{+6.00 \\ -4.56}$ & 14.87$\substack{+2.90 \\ -3.00}$ & 4.51$\substack{+0.11 \\ -0.11}$ & 0.49$\substack{+0.04 \\ -0.04}$ & 0.71$\substack{+0.03 \\ -0.03}$ & 0.84$\substack{+0.02 \\ -0.02}$\\
\vspace{0.8mm}
10\,000 deg$^{2}$ @ 10 h (with 25 per cent) & 32.10$\substack{+6.87 \\ -4.97}$ & 14.81$\substack{+2.91 \\ -3.01}$ & 4.52$\substack{+0.12 \\ -0.12}$ & 0.49$\substack{+ 0.05\\ -0.04}$ & 0.71$\substack{+0.03 \\ -0.03}$ & 0.84$\substack{+0.02 \\ -0.02}$\\
\hline
\end{tabular}
\caption{Summary of the median recovered values (and associated 16th and 84th percentile errors) for our three \eor{} model parameters, $\zeta$, $R_{\rm mfp}$ and $T^{\rm min}_{\rm vir}$ and the associated \igm{} neutral fraction, $\bar{x}_{\hi{}}$ for all considered observing strategies (with a 10 and 25 per cent modelling uncertainty and without a modelling uncertainty). Our fiducial parameter set is ($\zeta$, $R_{\rm mfp}$, ${\rm log}_{10}T^{\rm min}_{\rm vir}$) = (30, 15~Mpc, 4.48) which results in an \igm{} neutral fraction of $\bar{x}_{\hi{}} = 0.48$, 0.71, 0.83 at $z=8$, 9 and 10 respectively.}
\label{tab:3param_strat}
\end{table*}
 
Finally, it is important to note that given all three observing strategies are expected to be performed by SKA1--low, we can combine our constraints across all three surveys into one single measurement. In doing so, we find improved fractional uncertainties on our recover \eor{} parameters of 2.1 per cent for $\zeta$, 2.2 per cent for $R_{\rm mfp}$ and 0.3 per cent for ${\rm log}_{10}T^{\rm min}_{\rm vir}$. 

\subsection{Recovered \eor{} parameters including modelling uncertainty} \label{sec:wModUncert}

Our limits in the previous section assumed that we are able to perfectly simulate the EoR without any numerical errors. However, this is certainly not the case. Even large, numerical simulations differing only in their implementation of radiative transfer can result in notable differences for a fixed source model \citep[e.g.][]{Iliev:2006p3393}. Furthermore, the simplifications and approximations within semi-numerical approaches result in even larger modelling uncertainties, which appear to be of the order of 10's of per cent on scales relevant for recovery of \eor{} astrophysical parameters from the PS \citep[e.g.][]{Zahn:2011p1171}.

Therefore, in this section we present parameter forecasts in the presence of modelling uncertainties. Note, however, that this uncertainty only arises due to existing differences between current simulations and theory, and can therefore be reduced and/or statistically characterised in future work. For simplicity here we include this modelling uncertainty simply as a constant multiplicative factor of the modelled 21 cm PS, summing it in quadrature with the total noise PS (equation~\ref{eq:T+S}). Although in principle this should be a correlated error, this approach serves to showcase the relevant trends, before a more comprehensive statistical characterisation becomes available.

\begin{figure*} 
	\begin{center}
		\includegraphics[trim = 0.2cm 0.5cm 0cm 0.5cm, scale = 0.88]{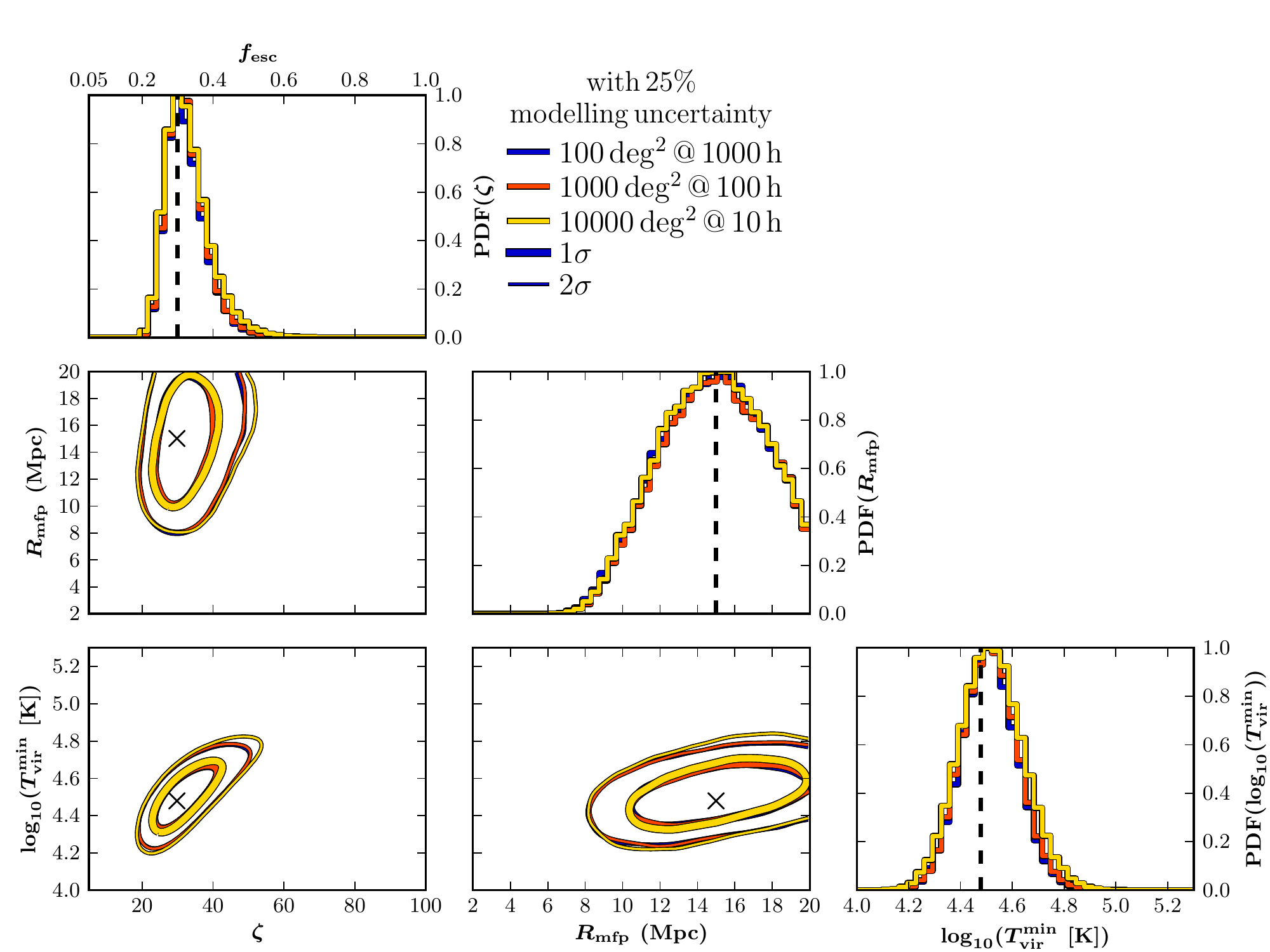}
	\end{center}
\caption[]
{Same as Fig.~\ref{fig:ObsStrat_woRT} except now we include a 25 per cent modelling uncertainty of the sampled 21 cm PS within \cmmc{}. This accounts for the expected fractional variation between the various numerical approximations adopted within the semi-numerical simulation code \cmfst{} relative to fully numerical reionization simulations.}
\label{fig:ObsStrat_wRT}
\end{figure*}

In Fig.~\ref{fig:ObsStrat_wRT}, we provide the same 1 and 2D marginalised likelihood contours from \cmmc{} for our \eor{} model parameters, but with the additional inclusion of a 25 per cent modelling uncertainty. We see a notable increase in all contour widths.  Moreover, all three strategies perform comparably. This highlights the need for parameter studies characterising modelling uncertainty, which is the limiting uncertainty for all of the SKA1-low surveys. We again caution that this only applies to the PS of the signal: additional information from non-Gaussian statistics (e.g. a simultaneous imaging analysis with the deep survey) could further improve on these constraints.

In the lower half of Table~\ref{tab:3param_strat} we quantitatively summarise the above behaviour. The fractional precision to which we are able to recover our \eor{} parameters for our 25 (10) per cent modelling uncertainty is 17.9/18.0/20.2 (7.6/8.2/10.4) per cent for $\zeta$, 18.5/18.2/18.4 (12.6/11.6/12.0) per cent for $R_{\rm mfp}$ and 2.4/2.4/2.7 (1.1/1.2/1.6) per cent for ${\rm log}_{10}T^{\rm min}_{\rm vir}$ for our deep, medium-deep and shallow observing strategies, respectively. We find both the deep (100 deg$^{2}$ at 1000~h) and medium-deep (1000 deg$^{2}$ at 100~h) surveys to be essentially equivalent for both our adopted choices of modelling uncertainty, with the shallow survey (1000 deg$^{2}$ at 100~h) performing marginally worse (as noted by an increase in errors for both $\zeta$ and ${\rm log}_{10}T^{\rm min}_{\rm vir}$).

The source of this shift in preference to deeper surveys is straightforward to understand. The fractional precision of the astrophysical constraints are strongly dependent on the sensitivity to which the large-scale modes are able to be sampled. By including a constant multiplicative error, we now substantially increase the errors on large scales, completely overwhelming any potential gains by shortening the observing times (decreasing the cosmic variance). In effect, all observing strategies now have the same sensitivity to the 21 cm PS on large scales. On smaller scales, the smaller thermal noise contribution from the deep and medium-deep surveys gives them the edge in parameter recovery over the shallow survey.
However, as noted in \GM{}, increased sensitivity on small scales does not significantly aid \eor{} constraints across multiple epoch observations, as the reionization history is still adequately sampled from the large-scales.

Finally, as in Section~\ref{sec:woModUncert}, with SKA1--low we can combine all three observing strategies to provide improved overall constraints on our \eor{} parameters. In the case of our 25 (10) per cent modelling uncertainty, we then find the improved constraints of 10.5 (4.9) per cent on $\zeta$, 10.7 (6.9) per cent on $R_{\rm mfp}$ and 1.4 (0.7) per cent on ${\rm log}_{10}T^{\rm min}_{\rm vir}$.

\section{Conclusion} \label{sec:conclusion}

The reionization epoch is rich in astrophysics: probing the formation and evolution of the first galaxies and their impact on the \igm{}.
Upcoming radio interferometers, such as SKA1--low and HERA, will provide a clear window into these processes, using the redshifted 21 cm line. It is therefore important to optimise their designs and observing strategies, so that we can maximise the corresponding EoR science returns.

Using the MCMC based \eor{} analysis tool \cmmc{} \citep{Greig:2015p3675}, here we explored the optimisation of SKA1--low.  Given the recent 50 per cent cost reduction, we explored how best to distribute the available resources, considering:
(i) halving the original number of core antenna stations (ii) halving the original number of dipoles per core station (maintaining the original number of antenna stations and doubling the field of view) and (iii) reducing the average baseline length by $\sqrt{2}$. We found the optimal core baseline design for SKA1--low to be one where the number of dipoles per station is halved with a concurrent reduction in baseline length. The former reduces the contribution from cosmic (sample) variance on large scales by observing a larger area of sky for a fixed survey duration, while the latter reduces the thermal noise on the relevant $k$-modes, resulting in the optimal sensitivity across all scales.

We then explored which planned observing strategy for SKA1--low provides the strongest constraints on our \eor{} model parameters from the 21 cm PS. SKA1--low plans to perform a three tiered observing approach \citep{Koopmans:2015:p1} (i) a single, deep 100 deg$^{2}$ survey area observed at 1000~h, (ii) a medium-deep survey of 1000 deg$^{2}$ at 100~h and (iii) a wide/shallow survey observing 10\,000 deg$^{2}$ at 10~h. We found the shallow survey to recover the tightest constraints on our model \eor{} parameters. By shortening the observation per field of view (large number of independent fields), the sample variance is considerably reduced, increasing the overall sensitivity on the largest scales which exhibit the largest differences between EoR models.  It is important to note that this conclusion is contrary to what would naively be expected from a simple consideration of the total S/N, which is maximised in the deep survey.  This highlights the need to use EoR parameter constraints as a figure of merit, instead of just the total integrated S/N.

However, once we included modelling uncertainty into our analysis, to account for approximations and simplifications inherent to our simulations, we found the strategies to be comparable, with the deep survey preforming slightly better than the others.
Modelling uncertainty dominates on large scales for all surveys, thus further motivating parameter studies which statistically characterise modelling error, in order to maximise the science returns from second generation interferometers.

It is worth emphasising that this optimisation study was based solely the 21 cm PS as a likelihood estimator. While this is the primary \eor{} science data product  of SKA1--low, other alternative statistics of the 21 cm \eor{} signal could yield additional information, especially in the deep survey where 21 cm structures can be directly imaged. These alternative statistics, to be explored in future work, will likely benefit from the increased sensitivities on smaller spacial scales.

Finally, we stress that in this study we only focused on the EoR.  The pre-reionization epoch of IGM heating encodes an additional wealth of information about high-energy processes inside the first galaxies.  For such Cosmic Dawn studies, deep integrations are needed to push out to the highest redshifts possible.

\section*{Acknowledgements}

This project has received funding from the European Research Council (ERC) under the European Union's Horizon 2020 research and innovation programme (grant agreement No 638809 -- AIDA). LVEK acknowledges the financial support from the ERC under ERC-Starting Grant FIRSTLIGHT -- 258942. We thank both Adrian Liu and Jonathan Pober for detailed discussions and insightful comments on a draft version of this manuscript.

\bibliography{Papers}

\end{document}